\documentclass[twocolumn,showpacs,superscriptaddress,preprintnumbers,amsmath,amssymb,prc]{revtex4}

\usepackage{graphicx}% Include figure files
\usepackage{dcolumn}% Align table columns on decimal point
\usepackage{bm}% bold math

\usepackage{color}
\usepackage{CJK}

\begin{document}
\begin{CJK*}{GBK}{song}

\title{Photonuclear reaction as a probe for $\alpha$-clustering nuclei  in quasi-deuteron region }
\author{B. S. Huang}
\affiliation{Shanghai Institute of Applied Physics, Chinese Academy of Sciences, Shanghai 201800, China}
\affiliation{University of Chinese Academy of Sciences, Beijing 100049, China}
\author{Y. G. Ma\footnote{Email: Corresponding author.  ygma@sinap.ac.cn}}
\affiliation{Shanghai Institute of Applied Physics, Chinese Academy of Sciences, Shanghai 201800, China}
\affiliation{ShanghaiTech University, Shanghai 200031, China}
\author{W. B. He}
\affiliation{Shanghai Institute of Applied Physics, Chinese Academy of Sciences, Shanghai 201800, China}
\affiliation{Institute of Modern Physics, Fudan University, Shanghai 200433, China}

\begin{abstract}
 Photon-nuclear reaction in a transport model frame, namely an Extended Quantum Molecular Dynamics (EQMD) model, has been realised at the photon energy of 70-140 MeV in the quasi-deuteron (QD) regime. For an important application, we pay a special focus on photonuclear reactions of $^{12}$C($\gamma$,np)$^{10}$B  where $^{12}$C is considered as different configurations including $\alpha$-clustering.  Obvious differences for some observables have been observed among different  configurations, which can be attributed to spatial-momentum correlation of a neutron-proton pair inside nucleus, and therefore it gives us a sensitive probe to distinguish the different configurations  including $\alpha$ clustering with the help of the photonuclear reaction mechanism.
\end{abstract}

%\pacs{25.20.-x, 21.60.Gx, 24.10.-i}

\maketitle

\section{Introduction}

Photonuclear reaction has been investigated  for several decades and it is considered as an important process for the understanding of the nuclear  structure and the fundamental dynamics of the nucleonic system for several reasons: 1) The availability of high-quality monochromatic photon beams generated by tagged photon technique or electron-laser Compton backscattering gamma sources ~\cite{Nuclear photonfissility,Duke,SIOM,ELI,SINAP, Amano}; 2)
It is helpful to investigate the behaviour of hadrons in nuclear medium, and 3) These probes are elementary and non-hadronic and  therefore allow us in principle to obtain information about the whole nuclear volume.
In the past decades, most researches focused mainly on low photon energy region, for an example, in the giant dipole resonance (GDR) region at photon energy of 15-40 MeV \cite{photo_16O}.  When the photon energy gets higher than the GDR region and  approximately reaches to 140 MeV, the wavelength of the photons is typically smaller than the size of the nucleus, which is close to the size of the deuteron. To deal with this region, the quasi-deuteron absorption mechanism has been introduced \cite{J.S.Levinger0}. It is indicated that the photonabsorption of a proton and neutron pair in the nucleus is dominated in this region, and therefore this process can provide a method for the study of the nucleon-nucleon (NN) correlation in nucleus.

As for target nucleus in photon-induced reactions, especially for $\alpha$-conjugate light nucleus, $\alpha$-clustering state plays one of the fundamental roles in current nuclear physic and nuclear astrophysics, which is crucial for the process of nuclear-synthesis and the abundance of elements ~\cite{Ikeda,Greiner,Ortzen,Freer,JBN,Nature,THSR}.
For the $\alpha$-clustering nucleus, the emergent properties are rich because of their different configurations and shapes \cite{Schuck,16O_chain,CFT1,CFT2,D3,Zhou,NST}. For instance, giant dipole resonance (GDR)  displays corresponding characteristic spectra for different configurations of $^{12}$C and $^{16}$O~\cite{W.B.He}.
Some aspects of $\alpha$-clustering behaviour have been discussed \cite{Ortzen}. Therefore, the $\alpha$-clustering nucleus is a good choice as a target nucleus for our photonuclear reactions. $^{12}$C is a good choice not only it  is of particular importance in its use as the standard from which atomic masses of all nuclides are measured but also an interesting three-$\alpha$ clustering nucleus involved in astrophysical nucleus-synthesis with its Hoyle state \cite{Hoyle}. A few years ago, the p-n correlation in $^{12}$C has been also studied by the two-nucleon knockout reaction \cite{np}.

Even though a variety of experimental and theoretical studies, there are only a few studies and discussion for the process of photonabsorption that considers for the effect of $\alpha$-clustering with different configurations in  target nuclei. For instance, photodisintegration of $^9$Be through the 1/2$^+$ state near neutron threshold and cluster dipole resonance below giant dipole resonance was measured with quasi-monochromatic $\gamma$-ray beams produced in the inverse Compton scattering of laser photons \cite{Utsno}.
In this work, we present a transport model calculation for photon-nuclear reaction of  a nucleus with the $\alpha$-clustering structure in the incident energy around 100 MeV. Using quasi-deuteron mechanism, we calculate the photon absorption of $^{12}$C  and demonstrate  the difference between different configurations.

\section{EQMD model }

Quantum molecular dynamics (QMD) type models~\cite{J.Aichelin,J.Aichelin2} have been extensively applied for dealing with fragment formation and nuclear multifragmentation in heavy ion collisions at intermediate energy successfully \cite{J.Aichelin2,C.Hartnack,C.Hartnack1}. It can also treats for the studies of giant dipole resonance (GDR), pygmy dipole resonance (PDR), and giant monopole resonance (GMR) \cite{C.Tao,C.Tao2,C.X.Guang,H. L. Wu,Ye,Ye2}. However, the description of the ground state of the nuclear system is not accurate enough in the usual QMD type model, because the phase space obtained from the samples of Monte Carlo is not in the lowest point of energy~\cite{MARUYAMA}. To compensate for this shortcoming, two features of the model are important. One is the capability to describe nuclear ground states, and the other is the stability of nuclei in the model description. The standard QMD shows insufficient stability due to the fact that the initialized nucleus is not in their real ground states. To solve this problem, an extended version of QMD called as EQMD which is used in our calculation has been developed ~\cite{MARUYAMA}.

Two features are introduced in EQMD in comparison with the standard QMD. In order to cancel the zero-point energy caused by the wave packet broadening in the standard QMD, the cooling process can be used to keep the mathematical ground state, but the Pauli principle is broken.
As the usual QMD model, Fermi
statistics is not satisfied in the present EQMD  because nucleons are not antisymmetrized. However,  repulsion between identical nucleons is
phenomenologically taken into account by a repulsive potential ~\cite{A.Ohnishi}, called as a Pauli potential.
As a result, saturation property and cluster structures can be obtained after energy cooling in the EQMD
model  \cite{W.B.He}.
 Another feature is that EQMD model treats the width of each wave packet as a dynamical variable~\cite{P.Valta}. The wave packet of nucleon is taken  as the form of Gaussian-like as
\begin{multline}
\phi_{i}(r_{i})=\bigg(\frac{v_i+v^{*}_{i}}{2\pi}\bigg)^{3/4}exp\bigg[-\frac{v_{i}}{2}(\vec{r}_{i}-\vec{R}_{i})^{2} +\frac{i}{\hbar}\vec{P}_{i}\cdot \vec{r}_{i}\bigg],
\end{multline}
where $\vec{R}_{i}$ and $\vec{P}_{i}$ are the centers of position and momentum of the $i$-th wave packet, and the $v_{i}$ is the width of wave packets which can be presented as
${v_i} = {{1/{\lambda _i}}} + i{\delta _i}$  where $\lambda_i$ and $\delta_i$ are dynamic variables.
The ${v_i}$ of Gaussian wave packet for each nucleon is dynamic and independent.

The Hamiltonian of the whole system is written as
\begin{multline}
H =\left\langle \Psi \mid \sum_{i} -\frac{h^{2}}{2m}\bigtriangledown^{2}_{i}-\widehat{T}_{c.m.}+\widehat{H}_{int} \mid\Psi \right\rangle\\
\\=\sum_{i}\bigg[\frac{{\vec{P}_i}^2}{2m}+\frac{3\hbar^{2}(1+\lambda^{2}_{i}\delta^{2}_{i} )}{4m\lambda_{i}} \bigg]-T_{c.m.}+H_{int},
\label{eq_H}
\end{multline}
where $T_{c.m.}$ is the zero-point center-of-mass kinetic energy, the form  can be found in details in Ref.~\cite{A.Ono}.

In Eq.~\ref{eq_H}, $H_{int}$  is the interaction  potential with the form of
\begin{equation}
H_{int} = H_{Skyrme} + H_{Coulomb} + H_{Symmetry} + H_{Pauli},
%\label{eq1}.
\end{equation}
where the Pauli potential $H_{Pauli}$ is written as
\begin{equation}
H_{Pauli}=\frac{c_{ P}}{2}\sum_{j}(f_{i}-f_{0})^{\mu}\theta(f_{i}-f_{0})
%\label{eq1}.
\end{equation}
with $f_{i}$ defined as an  overlap of $i$-th nucleon with other nucleons which have the same spin and isospin.

 In the present work,  we shall simulate the photonabsorption in the EQMD model with the obtained configurations for $^{12}$C .

%%%\\\\\\\\\\\\\\\\\\\\\\\\\\\\\\\\\\\\\\\\\\\\\\\\\\\\\\\\\\\\\\\\\\\\\\\\\\\\\\\\\\\\\\\\\\\\\\\\\\\\\\\\\\\\\\\\\\\\\\\\

\section{ The approach for process of photon-nuclear reaction }

In this section we will describe the methodology  for  photonuclear reaction within the EQMD model. In the considered energy region, we treat photonabsorption mechanism by the quasi-deuteron. There are two steps in the whole process. The first step is the absorption process. We consider that the photon is absorbed by a proton-neutron pair in one alpha-cluster of $\alpha$-conjugate nuclei which are taken from the cooling process of EQMD. Spherical configuration as well as two  different $\alpha$-clustering configurations for $^{12}$C are taken into account. In the second step, the nucleus gets excited after the absorption process and then goes into transport process to final state (see the details in the following).

\subsection{The quasi-deuteron absorption mechanism }

The present calculation was focused on the intermediate energies of photons about 70-140 MeV  where the two-nucleon absorption mechanism plays a dominant role for the wavelength close to the size of a deuteron. Since the photodisintegration reaction is predominant  as it is for the deuteron, n-n and p-p pairs do not contribute
in this mechanism, a n-p pair in the nucleus absorbs the photon much like the photodisintegration of the deuteron. The quasi-deuteron mechanism that was first introduced by Levinger~\cite{J.S.Levinger} {\it et al.}  considers the reminder of the nucleons as spectator besides the correlated proton-neutron pair
and its cross section reads
\begin{equation}
\sigma_{QD} =\frac{L}{A}NZ\sigma_{d}(E_{\gamma}).
\label{eq1}
\end{equation}
The factor of $L$ is the Levinger's factor that indicates the differences in density between the real deuteron and the nucleus.
In previous study, lots of experimental work for tagged photon has been measured, especially for the light target nuclei. For examples, Doran {\it et al.}  have measured the ($\gamma$,$^{4}$He) \cite{Doran} and McGeorge {\it et al.} have presented the $^{12}$C($\gamma$,2N) measurements \cite{J.C.McGeorge}.

The QD model has been employed to access the total photo-absorption cross section in heavy nuclei, it is based on the assumption that the incident photon is absorbed by a correlated neutron-proton pair inside the nucleus, leaving remaining nucleons as spectators. Such an assumption is enforced when one compares the relatively small wavelength of the incident photon with the nuclear dimensions. The QD cross section is proportional to the free deuteron photo-disintegration cross section. The photo-disintegration of nucleus was studied theoretically using the quasi-deuteron model by Levinger \cite{J.S.Levinger} and later by Futami and Miyazima  \cite{Futami}.

\begin{figure*}
\center
\includegraphics[scale=0.7]{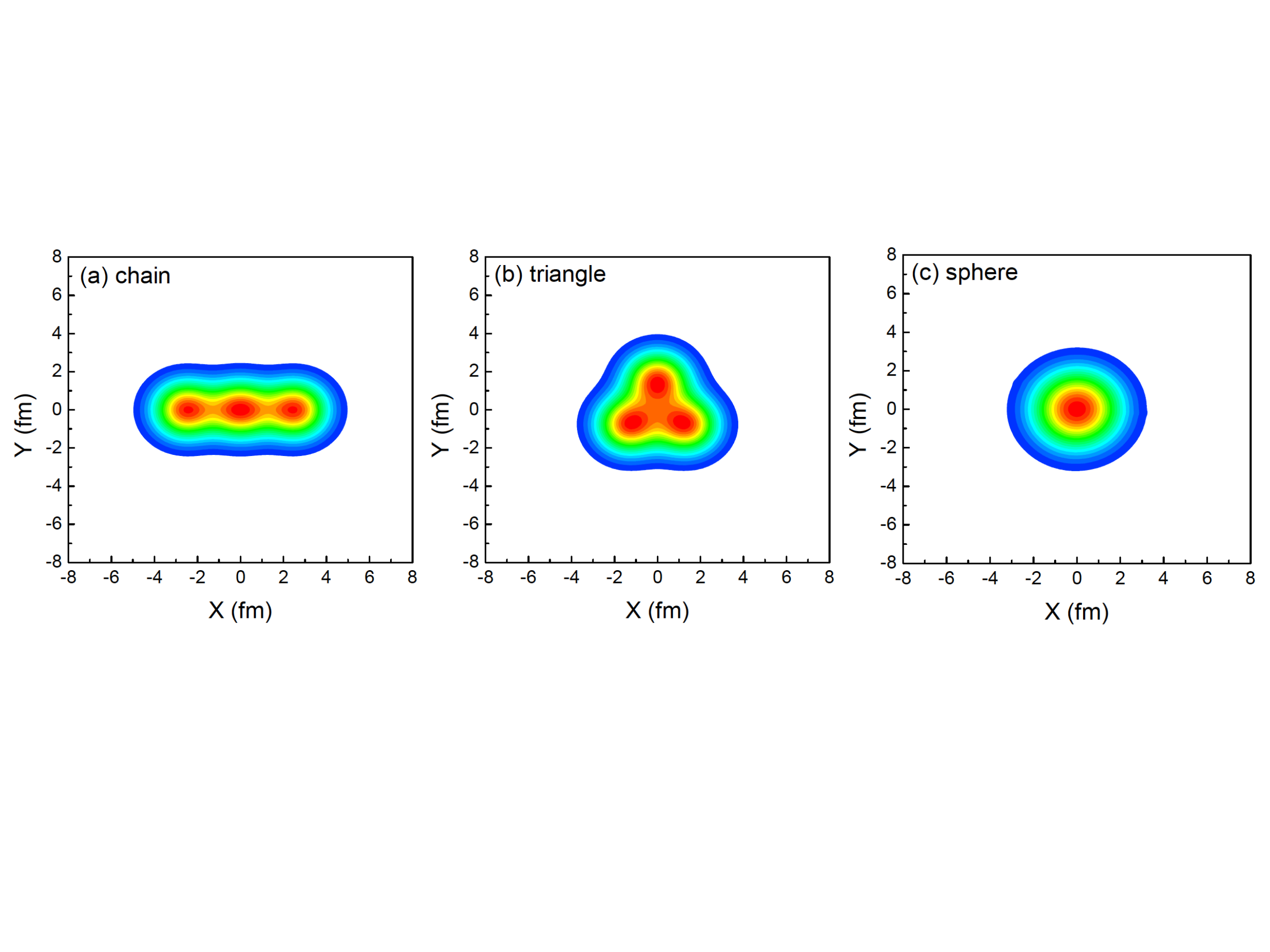}%new2.pdf}%JPG}
\vspace{-4cm}
\caption{Three different configurations of $^{12}$C were obtained in our EQMD model.
(a) chain 3-$\alpha$ structure;  (b) triangle  3-$\alpha$ structure;  (c) spherical structure without any $\alpha$-cluster. }
\label{Fig_cluster}
\end{figure*}

\subsection{Initial part of the process of photonabsorption}

Before we treat photonabsorption process, the reasonable initial phase space of target should be prepared. For traditional spherical structure of $^{12}$C, it can be easily obtained without an additional Pauli potential in the EQMD model.  Right panel of Fig.~\ref{Fig_cluster} shows the spherical $^{12}$C structure. For $\alpha$-clustering structures, we need to  introduce the Pauli potential
so that we can obtain $^{12}$C with two possible  three-$\alpha$ clustering structures as shown in  left and middle panels of Fig.~\ref{Fig_cluster}. To compare the ground state data of $^{12}$C which has a RMS radius of 2.47 fm \cite{NPA} and binding energy 7.68 MeV, our simulated data of three configurations are shown in Table ~\ref{table}.  In general, the bigger the binding energy, the nucleus tends to be more stable. Here, the spheric $^{12}$C is the most stable configuration, and the second is the chain structure and the third is the triangle structure. The reason why the chain structure seems stable than the triangle is due to the $\alpha$-cluster
at the center of $^{12}$C chain structure has a larger Gaussian wave
packet width, which can help to hold the $\alpha$ clusters at both
ends of the chain \cite{W.B.He2}.
From the viewpoint of RMS radius, the triangle structure is the best close to the ground state data, and the chain structure is the largest due to the hyper deformed shape. If we take the configuration with the largest binding energy as the ground state, our simulation gives the spherical shape, and then the triangle and chain 3-$\alpha$ structures represent the different excited states. In other literatures, three-alpha cluster models are also successful in describing
properties of ground and excited states of $^{12}$C and show that the
ground state has dominantly a triangle 3-$\alpha$ cluster structure.
Moreover, in microscopic cluster models, triangle 3-$\alpha$ cluster
structure is almost equivalent to a shell-model (oblate) configuration
because of the antisymmetrization. There, $\alpha$ clusters are largely
overlapping each other differently from the present picture of the
triangle 3-alpha configuration.

\begin{table}[]
\centering
\caption{ RMS radius and binding energy of different configurations of ${12}$C  and  the ground state data. }
\label{f0table}
\begin{tabular}{cccc}
\hline
      Configuration   ~~~       & $r_{RMS}$ (fm)  & ~~~$E_{bind}$ (MeV/nucleon)        \\ \hline
Chain &  2.71           & 7.17           \\
Triangle & 2.35 	          & 7.12          \\
Sphere &  2.23	           & 7.60        \\
Exp. Data &  2.47	           & 7.68        \\  \hline
\label{table}
 \end{tabular}
\end{table}

 For the following study on the process of photonabsorption, we use the above three different $^{12}$C configurations for the comparison.

Considering that the energy of incident photons  is selected ranges from 70 to 120 MeV in this work, where the QD effect is dominant, therefore, in the first step we only consider the $^{2}$H($\gamma$,np) process that photonabsorption occurs by a proton-neutron pair in the nucleus.
Considering that the distance between the $\alpha$ clusters in light nucleus is large in comparison with the size of quasi-deuteron, therefore,
the process of photonabsorption is assumed actually that the photons are absorbed by one of the $\alpha$-clusters in the
light nucleus in the initial process,
which is similar to the process of ($\gamma$,$^{4}$He), then we assume the rest of nucleons in this cluster and the reminder of clusters in nucleus as the spectators.
However,  in microscopic cluster models,  alpha clusters in triangle 3-alpha cluster
structure  are largely overlapping each other as mentioned above, which may make the above simple treatment of photonabsorption  in the initial process by one of the $\alpha$-clusters more complicate. Anyway, in current EQMD frame,
 we could accordingly replace this process of the photon absorbed by a neutron-proton pair within an $\alpha$-cluster by the reaction of $^{2}$H($\gamma$, np). The cross-section of $^{2}$H($\gamma$, np) is reflected by using the angular-dependent formulas of proton of this reaction,  fitted by Rossi {\it et al.} for photon energy range from 20 to 440 MeV in the CM frame ~\cite{P.Rossi}. The usual form and the fitted phenomenological function of the differential cross section is presented as
\begin{equation}
\frac{d\sigma}{d\Omega}=\sum_{i}A_{i}(E_{\gamma})P_{i}(cos\theta),
\label{eq1}
\end{equation}
where $\theta$ is the angle between the incoming photon and the outgoing proton in the c.m. system, $P_i(cos\theta)$ are the Legendre polynomials, and $A_i$ is the coefficients. Details can be found in Ref.~\cite{P.Rossi}.

For each photonuclear reaction event only one $\alpha$ cluster is interacted with the photon. Using the total cross section of the $^{2}$H$(\gamma, np)$ to determine which $\alpha$ cluster that interacts with the photon in each event by Monte Carlo sampling.

\subsection{The kinematics part of photonabsorption}

After choosing one of the $\alpha$-clusters by Monte Carlo sampling according to the cross section formula of $^{2}$H($\gamma$,np), we select one pair of the proton and neutron randomly within this $\alpha$ cluster. The total 4-momentum in the system for the photonabsorption in the lab frame can be written as
 \begin{equation}
 \vec{P}^{Lab}_{tot} = \vec{P}^{Lab}_{\gamma} + \vec{P}^{Lab}_{QD}.
\label{eq2}
\end{equation}
 Then we translate to the c.m. frame by the Lorentz boost. The total momentum of system before absorption is like this
\begin{equation}
 \vec{P}^{cm}_{tot} = L(\beta) \vec{P}^{Lab}_{tot},
 \label{eq2}
\end{equation}
where
$ \beta={P}^{LAB}_{tot}/P^{LAB}_{tot}(0)$, $L(\beta)$ is the operation of the Lorentz transformation, and $\vec{P}^{Lab}_{tot}(0)$ is the total energy of the two-body system in c.m. frame.

In term of conservation of momentum and energy, the 4-momentum of outing pair of nucleons of $^{4}$He($\gamma$, pn)d is  written as following
 \begin{eqnarray}
 E^{cm}_{p} &=&E^{cm}_{n} = P^{cm}_{tot}(0)/2, \\
\vec{P}^{cm}_{p} &=&-\vec{P}^{cm}_{n}=\sqrt{m^{2}+(\vec{P}^{cm}_{tot}(0)/2)^{2}},
\label{eq2}
\end{eqnarray}
 where the $m$ is mass of nucleon. The angular distribution of outgoing nucleons is obtained by the differential cross-section of $(\gamma,np)$  using a Monte Carlo sampling of the $^{2}$H($\gamma$,p)n differential cross-section (see Eq. \ref{eq1}). We assume that the incoming photons are randomly distributed in xy-plane, then we choose this event when the incoming photon was inside the region of QD total cross-section.
After the initial part for the process of $(\gamma,np)$ has been done,  the nucleus gets excited, and the nucleon could be emitted through final state interaction (FSI).

\section{ results and discussion}

In this section, we present several observables for  photodisintegration from different configurations of  $^{12}$C. For  $\alpha$-clustering configurations, their orientations are rotated randomly for each event. Even though there are different photodisintegration channels, here we only focus on one three-body channel, i.e. neutron, proton and a residue. Specifically for $^{12}$C, this residue is $^{10}$B.  Firstly we  discuss the recoil momentum and  missing energy  which can be compared with experimental data. Then we present  the pair momentum  of proton and neutron as well as the angular distribution between them. Finally we demonstrate the hyperangle  of the residue relative to the centre of mass the neutron and  proton as well as the hyper-radius of  three-body decay.

\subsection{Missing energy and recoil momentum}

\begin{figure}
\center
\includegraphics[scale=0.25]{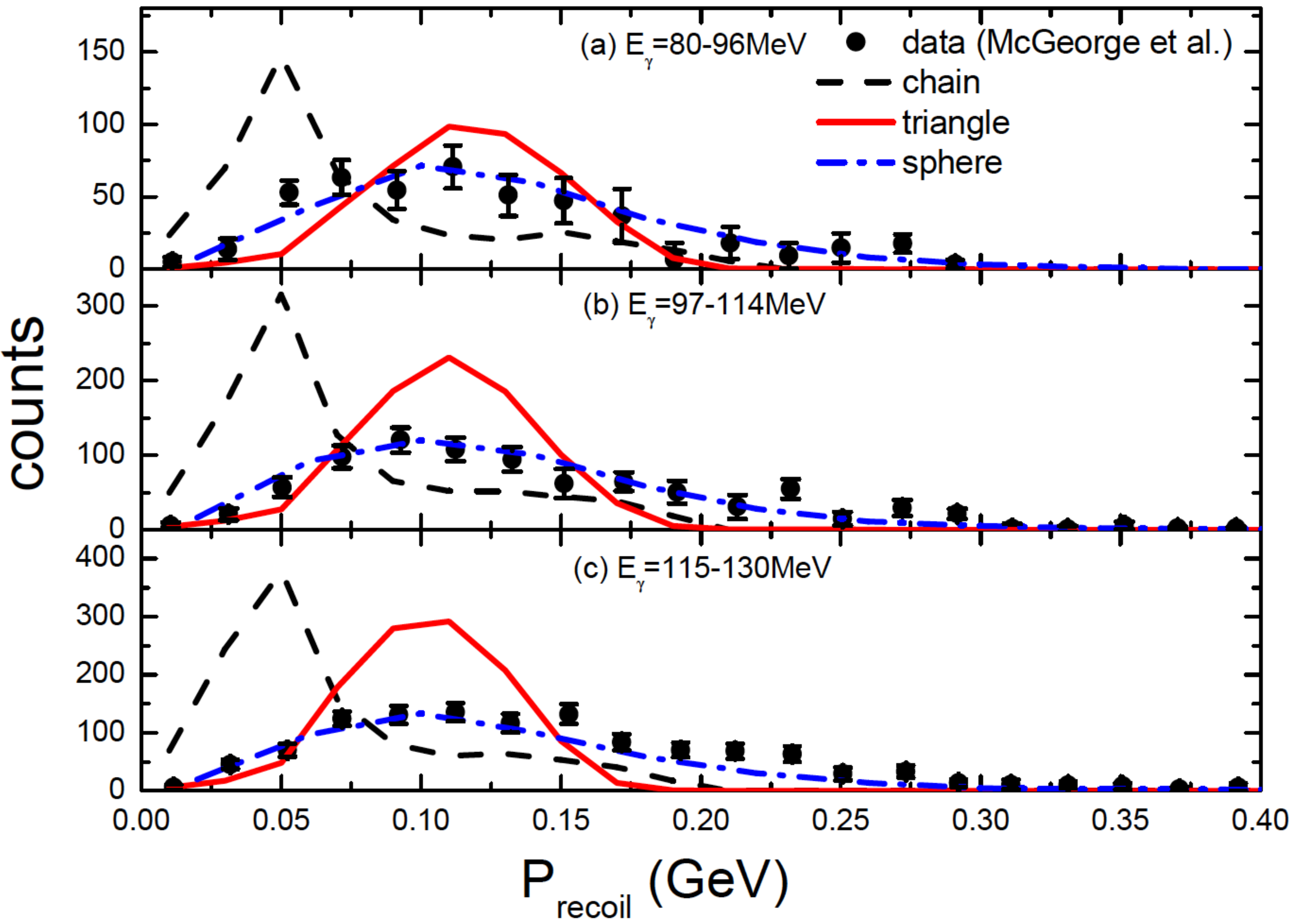}
\caption{The recoil momentum spectra of $^{12}$C with three configurations together  with the data (black dot)~\cite{J.C.McGeorge} in three incident photon energy widows. }
\label{fig_recol_p_12C}
\end{figure}

We calculate the recoil momentum and the missing energy  to compare with the experimental data. Using the distribution of bremsstrahlung with the weight of the 1/$E_{\gamma}$ , we obtain
the recoil momentum $\vec{p}_{recoil} = \vec{p}_{\gamma} - \vec{p}_{n} - \vec{p}_{p}$ event by event, where $\vec{p}_{\gamma}$ is the momentum of incident photon, $\vec{p}_{n}$ and $\vec{p}_{p}$ the momentum of emitted protons and neutrons, respectively. Fig.~\ref{fig_recol_p_12C} shows the recoil momentum spectra in three energy intervals of incident photons for three $^{12}$C configurations. Note that  the spherical result (blue dot dashed line) is normalised by the peak of the data (solid dots) and three different configuration results are normalised by the same events of the three body decay channel. From the top to low panel, one can see that the recoil momentum of the system for the chain $\alpha$-clustering structure is the smallest, and while the spherical and triangle $\alpha$-clustering structure is similar but with different width. From fits to the data, the spherical structure seems to have  the best fit.

\begin{figure}
\center
\includegraphics[scale=0.65]{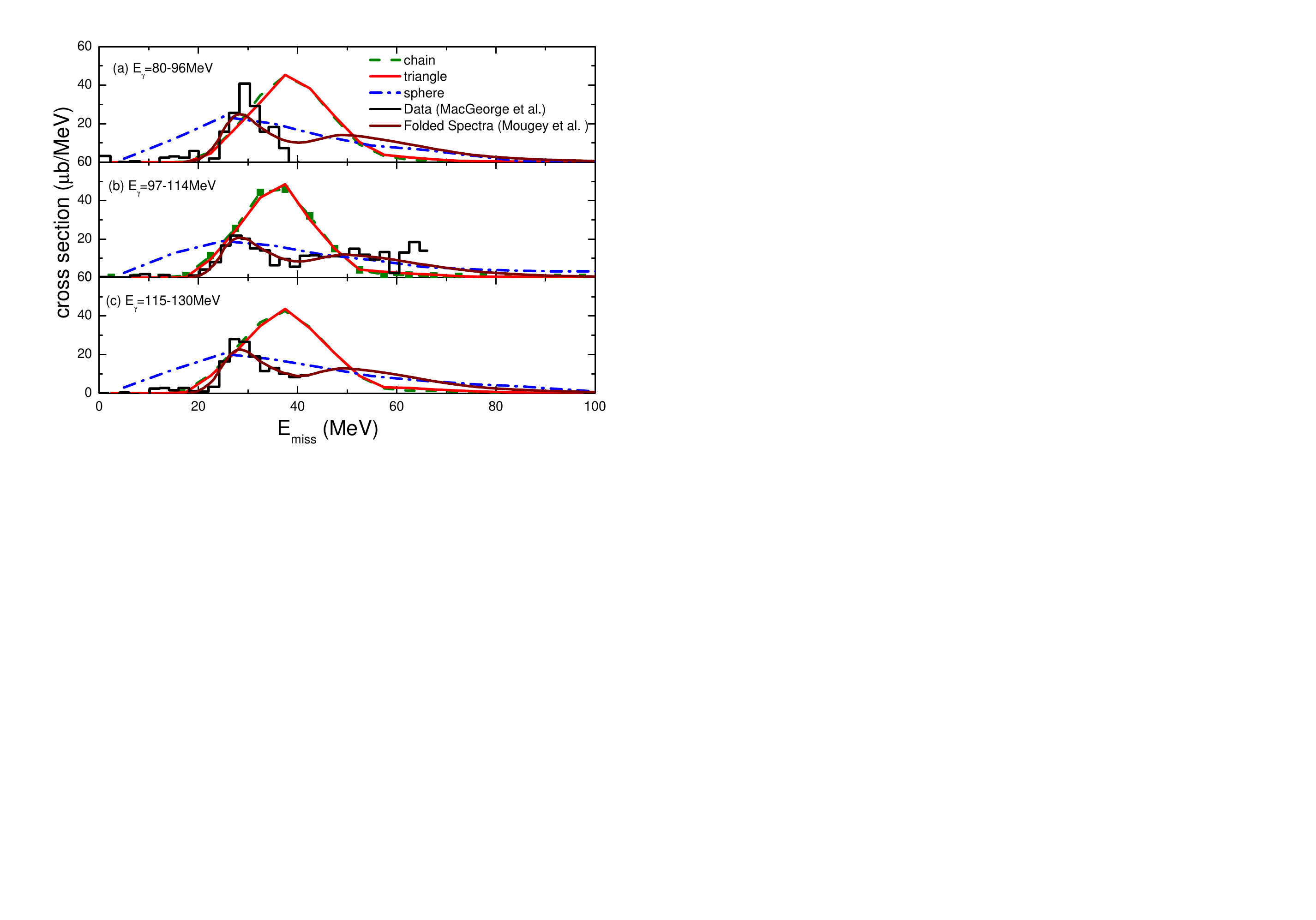}
\vspace{-7cm}
\caption{The missing energy spectra of $^{12}$C with with three configurations in different incident photon energy windows. The solid histograms represent the data and wine curves  are the results from the folding spectra derived from $^{12}C(e,e^{'}p)$ data. For details, see texts. }
\label{fig_Emiss}
\end{figure}

However, the shell structure effect could be also important, for instance, for the result of missing energy spectra ($E_{miss}$) for $^{12}$C which is  depicted in Fig.~\ref{fig_Emiss}.  Here   $E_{miss} = E_{\gamma} - T_{n} - T_{p} - T_{recoil}$, with $T_{n}$, $T_{p}$ and $T_{recoil}$ is defined as the kinetic energies of neutron, proton and the recoiled residue and $T_r$ was obtained from the recoil momentum  $\vec{p}_{recoil} = \vec{p}_{\gamma} - \vec{p}_{n} - \vec{p}_{p}$.
For the data~\cite{J.C.McGeorge} , solid histograms represent the results corrected for detector and threshold effects using the so called 2N Monte Carlo simulation  and the wine solid lines are the results from the folding spectra derived from $^{12}C(e,e^{'}p)$ data \cite{NEW}. Two peaks of the two histograms  refer to 1p and 1s1p shells, respectively~\cite{J.C.McGeorge}. Concerning the calculations,  the distributions are normalised by the same three-body events and peak values of spherical results are normalised by the cross section data around the peaks. The green dashed line and red solid line correspond to chain and triangle  $\alpha$-clustering configuration of  $^{12}$C calculated in EQMD, respectively, both give the similar $E_{miss}$ with the peak at around the mean value of two experimental peaks. And while the spherical $^{12}$C gives a very wide $E_{miss}$ with the peak position close to the main peak of the data. From these comparisons, no one can reproduce the data perfectly which indicates our model is not able to treat the fine structure effect.
In addition, from the comparison of $P_{recoil}$ and $E_{miss}$, it indicates that $P_{recoil}$  is a more sensitive probe for different  configurations  including $\alpha$-clustering.

\subsection{Pair momentum and opening angles of the emitted pair of neutron and proton}

\begin{figure}
\center
\includegraphics[scale=0.65]{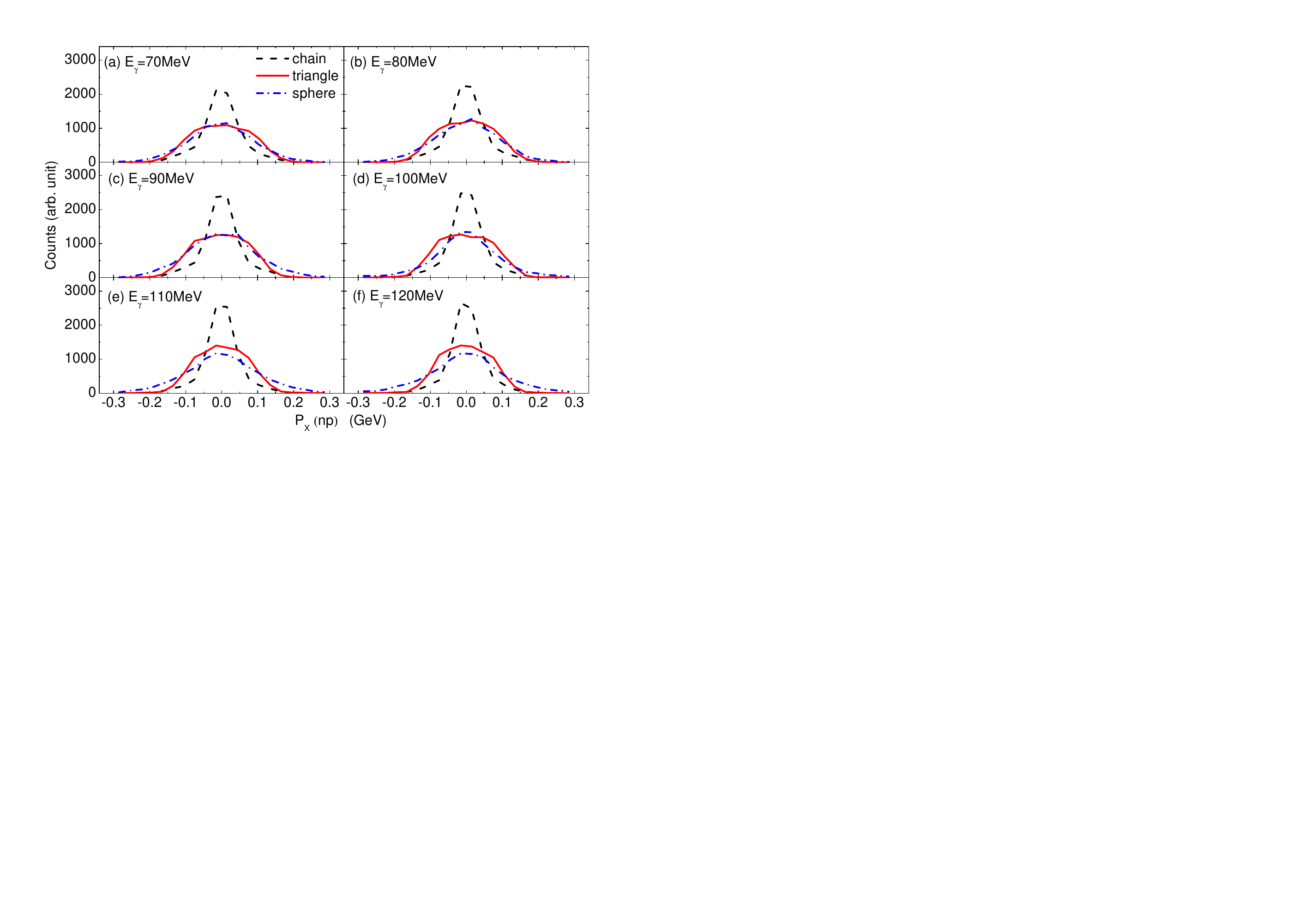}
\vspace{-7.5cm}
\caption{The pair momentum of  emitted proton and neutron from three configurations of $^{12}$C in six photon energy windows. }
%\vspace{-5.cm}
\label{fig_pari_p_12C}
\end{figure}

The sum of momentum of emitted proton and neutron pair along $p_x$ direction has been calculated. Fig.~\ref{fig_pari_p_12C}
plots  the pair momentum of the emitting proton and neutron in Px-axis direction  for three different configuration of $^{12}$C
in different incident photon energy windows.
From those panels, we can see that the width of the pair momentum is very distinct among different configurations. The general trend is that  the width of the chain structure is the narrowest, and  the width of spherical structure is the widest but close to the triangle structure.
This might be understood by the secondary scattering effect for the initial ejected neutron and/or proton. From the chain structure, the second scattering probability is obvious smaller due to its geometric structure. For the triangle and spherical structures, when the initial ejected neutron and/or proton passes through the remainder, it has higher probability to collide with the other nucleons, which will certainly increase the momentum width.
Furthermore, one can also observe that the trend for all width of the pair momentum  of $^{12}$C
basically remains unchanged even though the incident photon energy is different, which implies that their differences are mainly caused by geometric  effects.

\subsection{Angular distribution between the emitted neutron and proton}

\begin{figure}
\center
\includegraphics[scale=0.65]{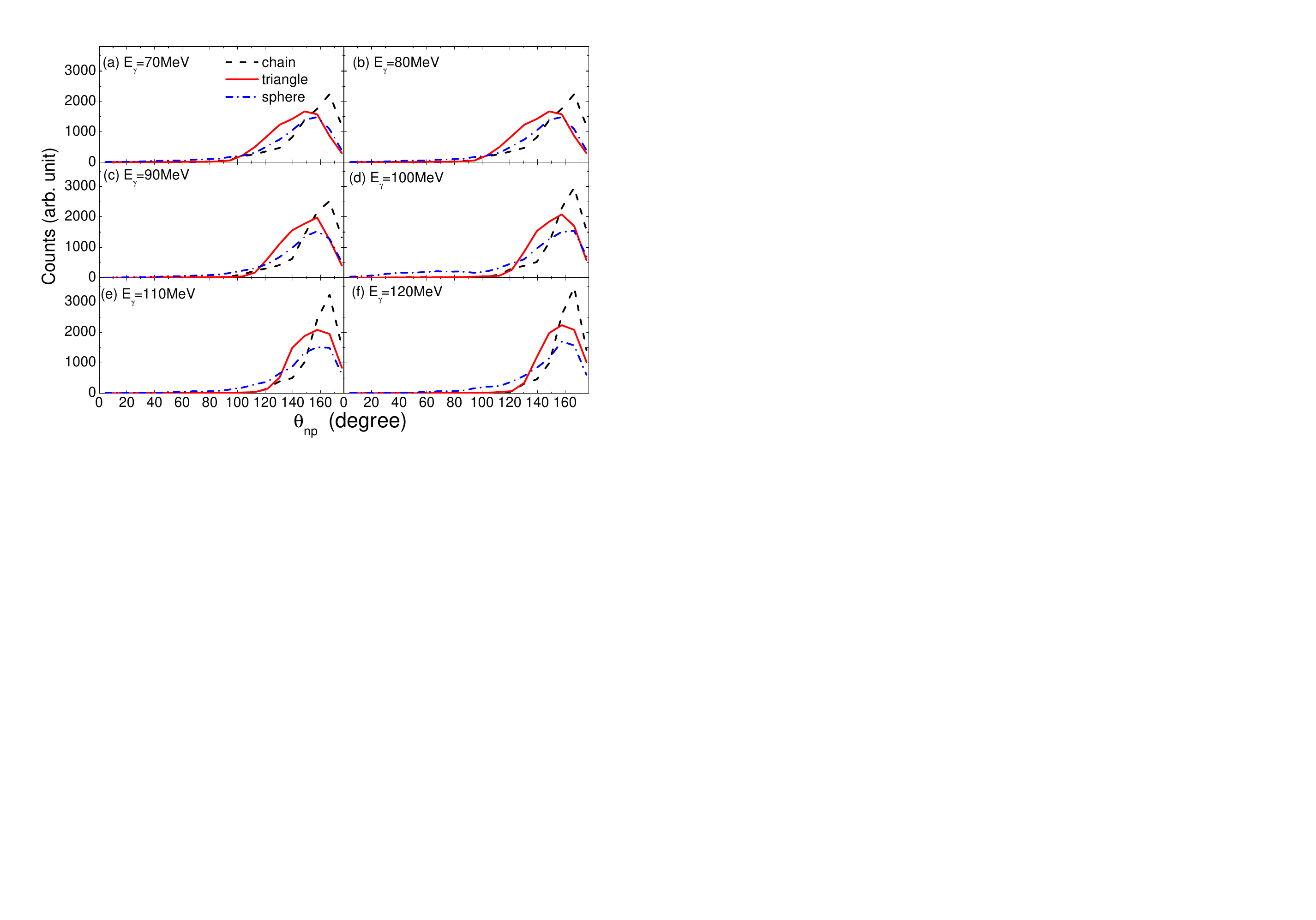}
\vspace{-7.2cm}
\caption{Same as Fig.~\ref{fig_pari_p_12C} but for the opening angular distribution between the emitted neutron and proton.}
\label{fig_theta_np_12C}
\end{figure}

Except the pair momentum of neutron and proton, we can investigate the opening angle between neutron and proton.  Fig.~\ref{fig_theta_np_12C}    presents the $\theta_{np}$-angular distribution between the emitted neutron and proton. One can see that the chain structure tends to back-to-back emission and while the triangle and spherical structures show a decreasing angle with wider distribution.
Again, for the chain configuration, the emitted neutron-proton pair which is almost back to back  in the initial state of the process of absorption will suffer less second collision with the residual nucleus in comparison with the triangle and spherical  configurations due to its linear shape  in space.
With the increase of the photon energy, the angular distribution seems a slight narrower, which can be understood by the less dissipation collision for the emitted neutron and/or proton with others in higher incident photon energy.

\subsection{Hyperangle and hyperradius of the three body decay }

Since we are treating three-body decay problem by the photonuclear reaction, the hyper-spherical formalism can be used. Here we  consider the emitted proton, neutron and the residual  nucleus as a three-body,  its $i$-th set of Jacobi coordinate $(x_i,y_i)$ is defined as ~\cite{E.Braaten,J.S.Levinger0,A.Tumio}
 \begin{eqnarray}
x_i &=& \mu_{jk}({\bf p_j} - \bf{p_k}), \\
y_i &=&  \mu_{i,jk} ({\bf p_i} - \frac{m_j {\bf p_j} + m_k {\bf p_k}}{m_j + m_k}),
\label{eq_x_y}
 \end{eqnarray}
 where
 \begin{eqnarray}
 \mu_{jk}&=&\sqrt{\frac{m_j m_k}{m(m_j + m_k)}},\\
\mu_{i,jk}&=&\sqrt{\frac{m_{i}(m_j + m_k)}{m(m_i + m_j + m_k)}},\\
\label{eq3}
 \end{eqnarray}
with $p_{j}$ and $p_{k}$ represent the momentum of emitted proton and neutron, $m_i$, $m_j$ and $m_k$ represent for the mass number of the residue nucleus, proton and neutron, respectively, and $m$ is the total mass number of the mother nucleus, i.e. $^{12}$C. The space-fixed hyper spherical coordinates can be expressed by
\begin{equation}
x_i = \rho sin(\alpha_i), ~~~~y_i = \rho cos(\alpha_i)
\label{Eq_hyperradisu}
\end{equation}
where $x_i$ and $y_i$ are the Jacobi momenta, $\rho$ is hyperradius, and $\alpha_i$ represents the hyperangle. If we assign $i$ as the index of the residue, $j$ and $k$ for neutron and proton, respectively, then $\alpha_i$ means the hyperangle of the residue to the neutron and proton.  Usually the hyperangle is confined by 0 $\leq  \alpha_{i}  \leq \frac{\pi}{2}$. If it is near 0, it indicates that the residual nucleus (i) is far from the proton (j) and neutron (k); If it is near to $\frac{\pi}{2}$, it indicates that the residue nucleus ($^{10}$B) is near  the center-of-mass of the emitted proton and neutron.

\begin{figure}
\center
\includegraphics[scale=0.65]{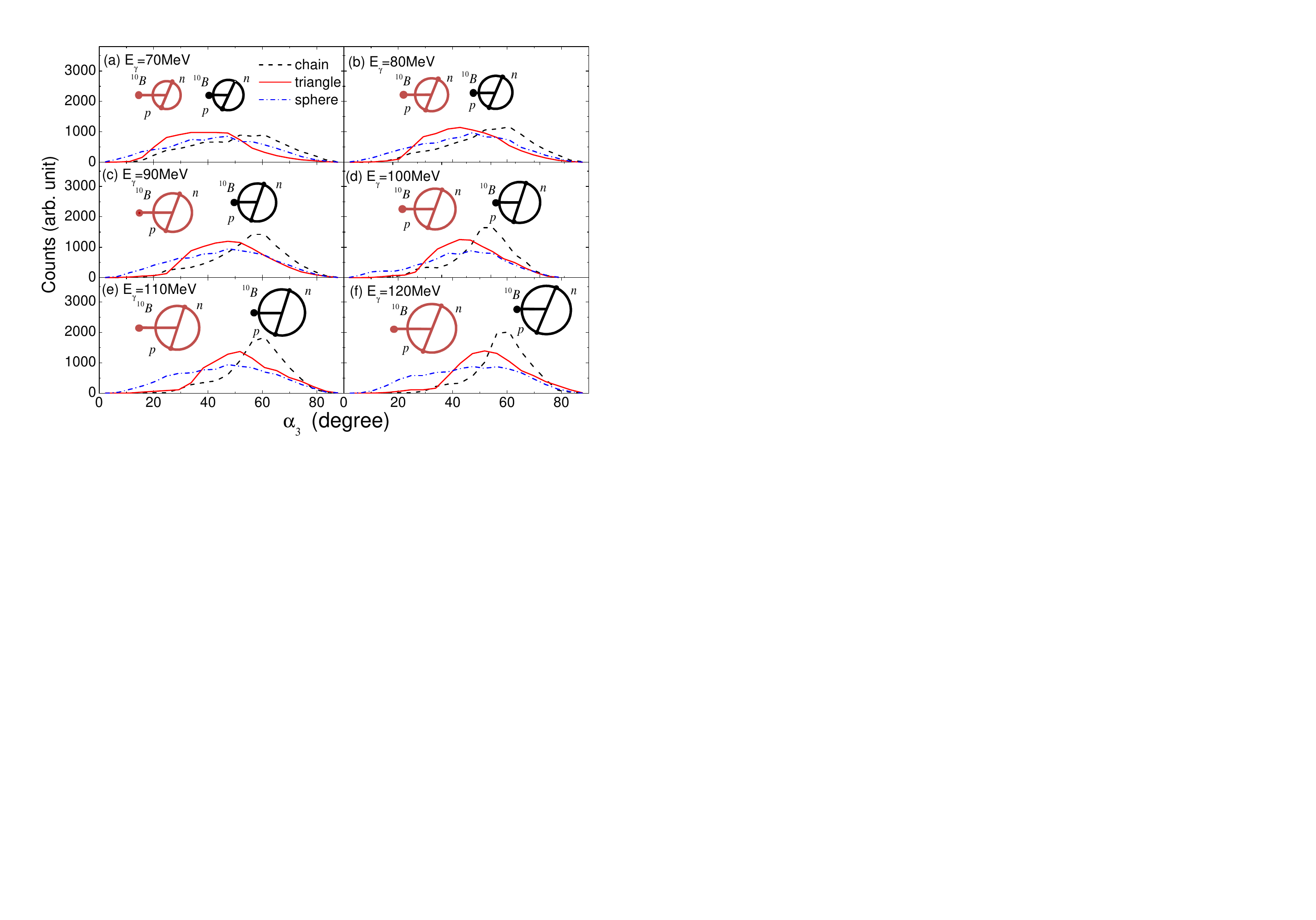}
\vspace{-7.2cm}
\caption{Hyperangular distribution  ($\alpha_{3}$) of  the residual nucleus $^{10}$B relative to the c.m. of neutron and proton from three-body decay of $^{12}$C for three configurations. Inserts display the schematic plots for the cases of chain 3-$\alpha$ structure (black) and triangle 3-$\alpha$ structure (red), respectively.}
\label{fig_hyperangle_12C}
\end{figure}

Fig.~\ref{fig_hyperangle_12C} presents the hyperangles ($\alpha_{3}$) of  the residual nucleus$^{10}$B relative to the c.m. of neutron and proton  for the $^{12}$C($\gamma$,np) with three $^{12}$C configurations. Generally the hyperangle of the chain $^{12}$C structure is relatively close to  $\frac{\pi}{2}$, indicating the residue $^{10}$B is  close to the center-of-mass of proton and neutron. And the spherical structure displays the widest distribution and the triangle 3-$\alpha$ structure is in between.
In the viewpoint of incident energy of photons, the distribution becomes narrower when the energy becomes higher, indicating more focusing effect for higher energy photons. In the figure, we also plot the schematic plots for the corresponding hyper-angle distribution. The black and red plots correspond to the chain and the triangle 3-$\alpha$ structure, respectively. Generally, the $^{10}$B is close to the c.m. of neutron and proton for the chain structure.

\begin{figure}
\center
\includegraphics[scale=0.65]{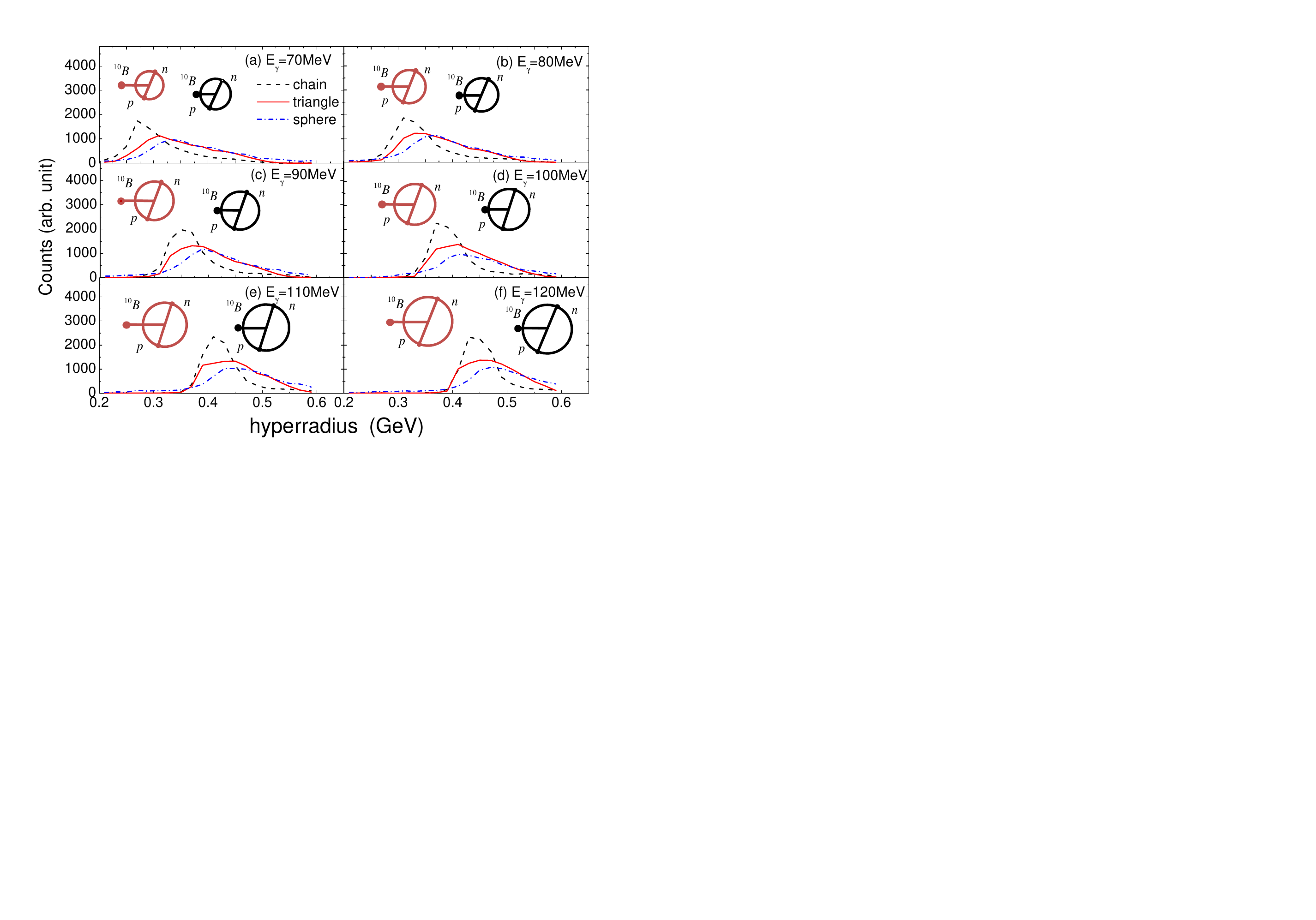}
\vspace{-7.2cm}
\caption{Same as Fig.~\ref{fig_hyperangle_12C} but for hyperradius distribution. }
\vspace{-0.5cm}
\label{fig_hyperradius_12C}
\end{figure}

Except hyperangle, hyperradius is another important quantity to characterise the property of the three-body decay.   The hyperradius is the root-mean-square separation of the three bodies, i.e. neutron, proton and the residue in the present work. The value of hyperradius is the $\rho$ in Eq.~\ref{Eq_hyperradisu}.
The hyperradius is small only if all three bodies are close together. It is large if any single product is far from the other two. From Figure \ref{fig_hyperradius_12C}, we can see the chain structure has smallest and narrowest distribution among three configurations.  It indicates that the decayed three bodies, namely the residue, neutron and proton are close together for the chain  structure, which is consistent with the above hyperangle distributions. While with the increase of photon energy,  hyperradius becomes larger, indicating that the  residue $^{10}$B   becomes more separated from neutron and proton for high energy photon reactions.  Same as Fig.~\ref{fig_hyperangle_12C}, the schematic plots are inserted for the chain structure (black ones) and the triangle structure (red ones).

\section{Summary}

In a framework of EQMD, we realisied, for the first time,  the photonuclear reaction  in the quasi-deuteron region. As an important application, we use the model to  investigate the photon's response for different configurations of $^{12}$C including  with $\alpha$-clustering structures at $E_\gamma$ = 70 - 140 MeV.
Firstly, the recoil momentum  and missing energy  for the calculations seem  consistent with experimental data in some extents, especially for the spherical case in the present model, which indicates that our  photonuclear reaction model based on the EQMD seems  reasonable. However, the present result does not exclude a
configuration with largely overlapping 3-$\alpha$ clusters that is
usually obtained by microscopic cluster models. With that strong overlapping 3-$\alpha$ clustering configuration, many properties will show up as spherical case  even though the structure could be totally different.
Further,  we calculate the pair momentum of the emitted neutron and proton as well as their angular distribution,
and find that  the chain structure has the narrowest pair momentum distribution and near back-to-back emission, and while the triangle three-$\alpha$ and spherical $^{12}$C distribution show the wider distribution.
From the hyperangle of the residue  relative to the c.m. of neutron and proton,
chain structure shows larger values than the triangular and spherical structures, which indicates that the residue from the chain structure is close to the center-of-mass of emitted proton and neutron.
In addition, the hyperradius results also display the smallest values for the chain structure and then illustrate that the emitted three bodies are much close together, which is consistent with the results of hyperangle. The above observables demonstrate  that the differences of pair momentum and angular distribution of emitted proton and neutron as well as three-body hyperangle and hyperradius  among different $^{12}$C configuration are sensitive to  structure of $^{12}$C, therefore offering a robust probe for  $\alpha$-clustering inside nucleus.

\vspace{2.cm}
\begin{acknowledgments}
This work was supported in part by the
National Natural Science Foundation of China under contract Nos. 11421505
and 11220101005, the Major State Basic Research
Development Program in China under Contract No. 2014CB845401, and the Strategic Priority Research Program of the Chinese Academy of Sciences (Grant No. XDB16).
\end{acknowledgments}

%%%%%%%%%%%%%%%%%%%%
%-------------------------------------------

\end{CJK*}

\end{document}